\documentclass{desyproc}

\usepackage{textcomp}
\usepackage{xspace}

\newcommand{\jpsi}{$\text{J}/\psi$\xspace}

\newcommand{\psitwos}{$\psi\left(2\text{S}\right)$\xspace}
\newcommand{\chic}{$\chi_c$\xspace}
\newcommand{\ipb}{pb$^{-1}$\xspace}
\newcommand{\sqrts}{$\sqrt{s}=7$~TeV\xspace}
\newcommand{\pt}{$p_{\perp}$\xspace}

\newcommand{\gevcc}{$\text{GeV}/c^2$\xspace}
\newcommand{\mevc}{$\text{MeV}/c$\xspace}
\newcommand{\mevcc}{$\text{MeV}/c^2$\xspace}
\newcommand{\um}{\textmu m\xspace}

\begin{document}

\title{Central exclusive J/$\boldsymbol{\psi}$ and $\boldsymbol{\chi_c}$ production at LHCb}

\author{{\slshape Scott Stevenson, on behalf of the LHCb collaboration}\\[1ex]
  Department of Physics, University of Oxford, Oxford, United Kingdom}

\contribID{44}

\acronym{EDS'13}

\maketitle

\begin{abstract}
  Central exclusive \jpsi and \chic meson production has been measured in
  decays to dimuons with the LHCb detector, in data corresponding to 36~\ipb of
  integrated luminosity from \sqrts proton-proton collisions. Cross-section
  measurements for \jpsi, \psitwos and $\chi_{c0,c1,c2}$ production are
  presented, and the \jpsi photoproduction cross-section is measured as a
  function of the photon-proton centre-of-mass energy and compared to
  measurements made at HERA.
\end{abstract}

\section{Introduction}

Central exclusive production in proton-proton collisions is a process in which
the initial state protons scatter elastically via colour singlet exchange. The
protons remain intact and escape undetected down the beampipe. Additional
particles are produced in the central region, accompanied by rapidity gaps.
This provides a clean experimental environment for measurements of the quantum
numbers of the produced state, and those processes involving pomeron exchange
provide a testing ground for the predictions of QCD. Dimuon production via
charmonium resonances has a high predicted cross-section and a clear
experimental signature and can be used to test the theoretical
predictions~\cite{SuperCHIC}. These proceedings present measurements of the
cross-section times branching fractions for exclusive \jpsi and \chic mesons to
produce two muons in the pseudorapidity range $2 < \eta < 5$ at a proton-proton
centre-of-mass energy of \sqrts. The measurements are compared to results from
HERA and a number of theoretical models.

\section{The LHCb detector}

The LHCb detector~\cite{JINST2008} is a single-arm forward spectrometer
covering the pseudorapidity range $2<\eta <5$. The detector includes a
high-precision tracking system consisting of a silicon-strip vertex detector
surrounding the interaction region, a large-area silicon-strip detector located
upstream of a dipole magnet, and three stations of silicon-strip detectors and
straw drift tubes placed downstream. The vertex detector allows reconstruction
of backward tracks in the range $-4 < \eta < -1.5$. Different types of charged
hadrons are distinguished by information from two ring-imaging Cherenkov
detectors.

Photon, electron and hadron candidates are identified by a calorimeter system
consisting of scintillating pad (SPD) and preshower detectors, an
electromagnetic calorimeter and a hadronic calorimeter. The scintillating pad
detector provides a measure of charged particle multiplicity. Muons are
identified by a system composed of alternating layers of iron and multiwire
proportional chambers.

The trigger consists of a hardware stage, based on information from the
calorimeter and muon systems, followed by a software stage, which applies a
full event reconstruction.

The unique forward acceptance of the LHCb detector, in addition to the low
pileup encountered and its sensitivity to low momentum particles, make it well
suited to central exclusive production studies.

\section{Event selection}

\jpsi and \chic mesons are reconstructed from their decays to
dimuons~\cite{JPHYSG2013,LHCb-CONF-2011-022}. The analyses are based on data
corresponding to an integrated luminosity of 36~\ipb collected at \sqrts in
2010. The hardware trigger requires a single muon with transverse momentum
$p_{\perp} > 400$~\mevc, or a dimuon, each track with $p_{\perp} > 80$~\mevc,
and less than 20 hits in the SPD. The software trigger requires a dimuon mass
greater than 2.9~\gevcc, or greater than 1~\gevcc with $p_{\perp} < 900$~\mevc
and a distance of closest approach of the tracks less than 150~\um. In the
offline selection the tracks are required to lie within $2.0 < \eta < 4.5$ and
the dimuon mass to be within 65~\mevcc of the \jpsi or \psitwos masses
\cite{PDG2012}. It is required there be no photons or other tracks (including
backwards tracks) in the detector, but for a single photon with $E_{\perp} >
200$~MeV in the \chic selection.

\section{Exclusive purity determination}

Three backgrounds are considered: non-resonant production via diphoton fusion,
exclusive \psitwos and \chic feed-down to the \jpsi sample, and inelastic
production with additional gluon radiation or proton dissociation.

\begin{figure}
  \centerline{\hfill\includegraphics[width=0.35\textwidth]{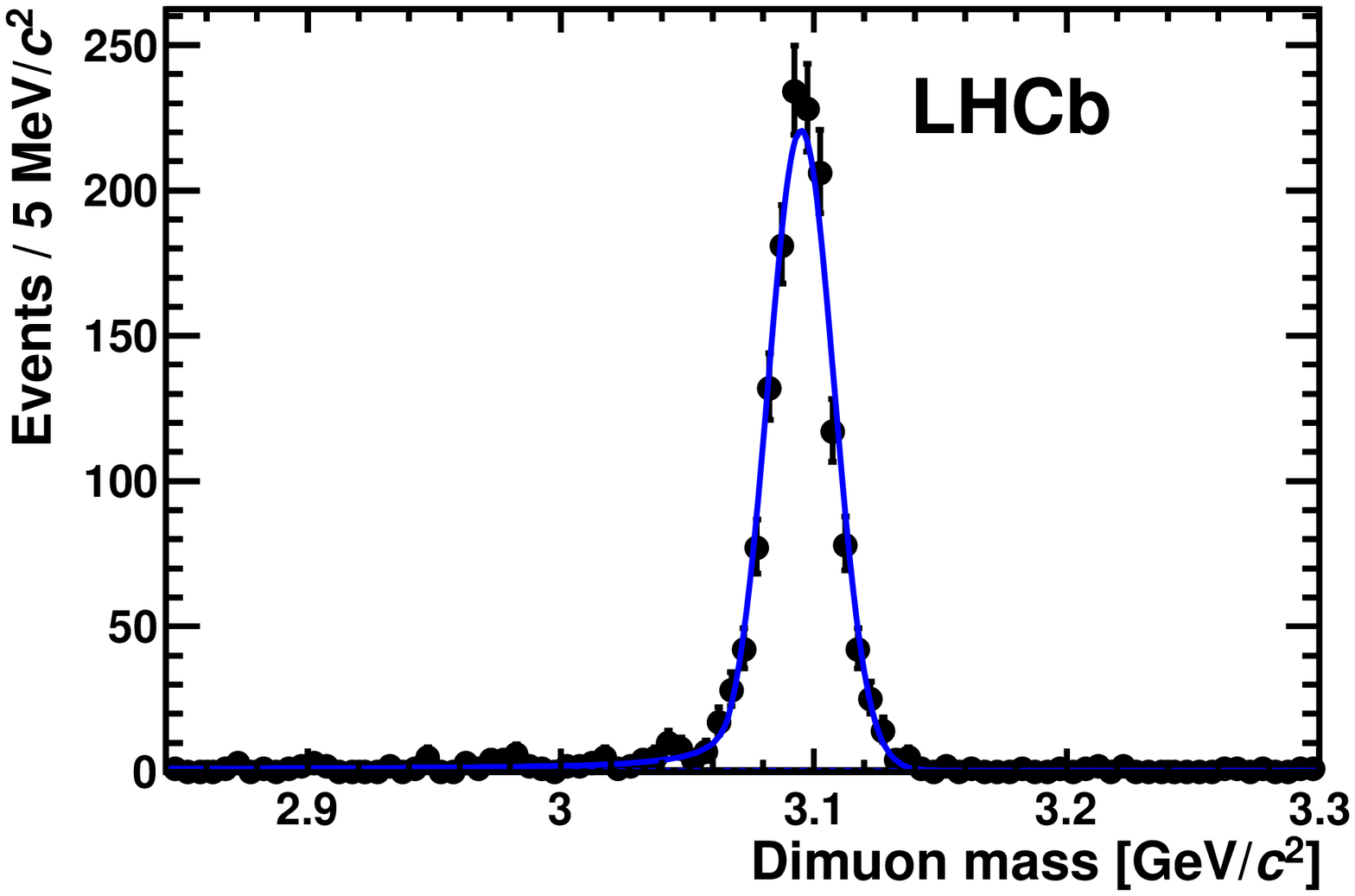}
    \hfill
    \includegraphics[width=0.35\textwidth]{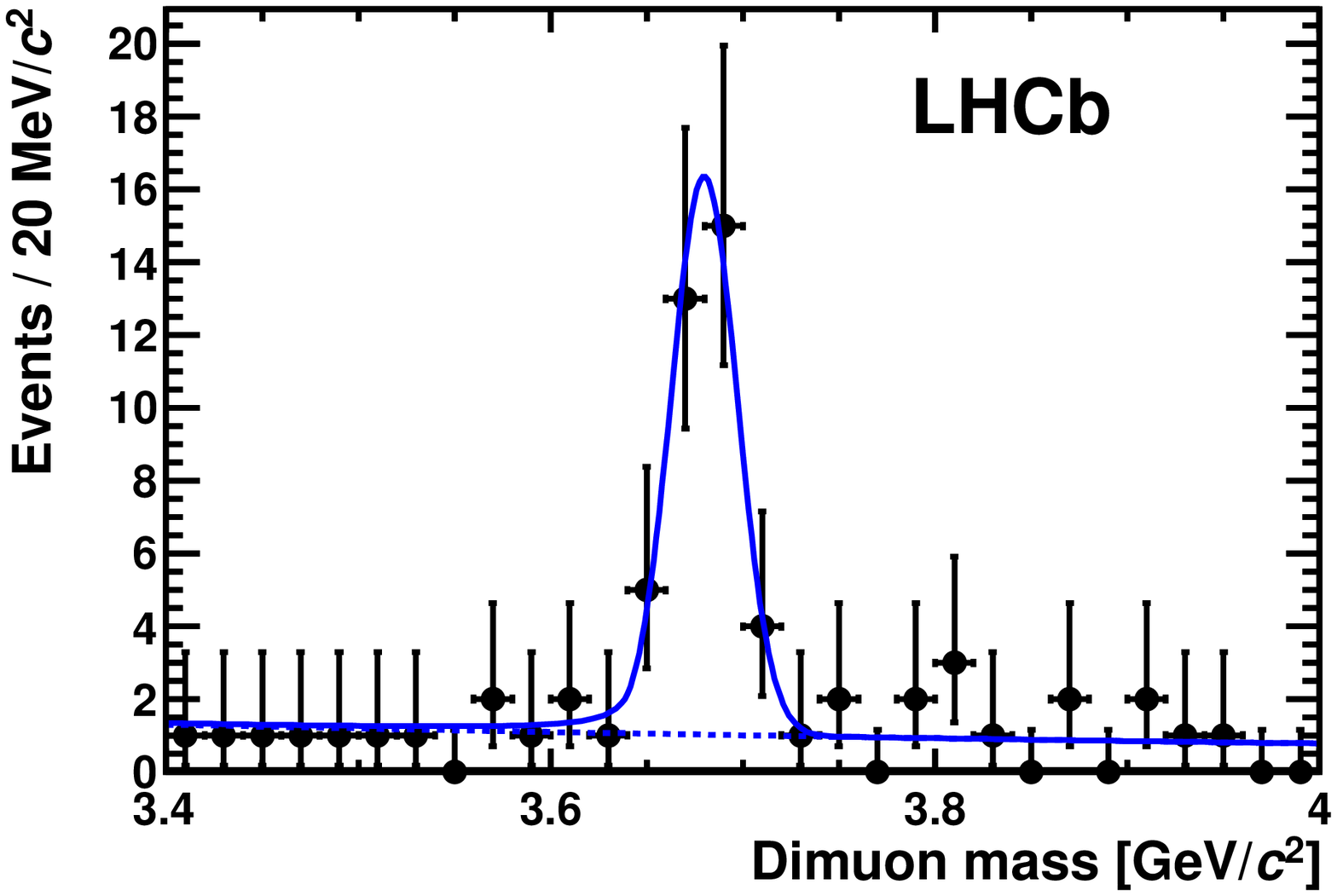}\hfill}
  \caption{Invariant mass distributions around (left) the \jpsi mass and
    (right) \psitwos mass for dimuons with $p_{\perp} < 900$~\mevc, in events
    with exactly two tracks and no photons. The dashed line shows the
    non-resonant contribution and the solid line the total
    fit.}\label{fig:mass_spectra}
\end{figure}

The non-resonant contribution is estimated by fitting the dimuon invariant mass
distribution, shown in Figure~\ref{fig:mass_spectra}. The resonances are
modelled with a crystal ball function and the continuum with an exponential.
The non-resonant contribution is estimated as $\left( 0.8 \pm 0.1 \right)$\%
and $\left( 16 \pm 3 \right)$\% of events within 65~\mevcc of the \jpsi mass
and \psitwos mass, respectively.

Exclusive \chic production can feed down to the \jpsi sample via $\chi_c
\rightarrow \text{J}/\psi\gamma$ where the photon is undetected. Events in data
containing a \jpsi and photon are identified as feed-down candidates and their
contribution estimated by scaling the \chic yield in data by the ratio of fake
exclusive \jpsi to exclusive \chic in simulation. The contribution is estimated
as $\left(9.0 \pm 0.8\right)$\%.

Exclusive \psitwos production can feed-down to the \jpsi sample via decays of
the type $\psi \left( 2 \text{S} \right) \rightarrow \text{J}/\psi X$ where
additional particles $X$ are undetected. The contribution is estimated from a
simulated \psitwos sample normalised to the $\psi \left( 2 \text{S} \right)
\rightarrow \mu \mu$ yield in data, and determined to be $\left(1.8 \pm
0.3\right)$\%.

The inelastic contribution is extracted from a fit to the dimuon \pt spectrum.
A Novosibirsk function fitted to events with three to eight tracks is
extrapolated to events with exactly two tracks for the background shape. The
signal shape is taken from simulation. The contribution for events with a
dimuon $p_{\perp} < 900$~\mevc is estimated as $\left(30 \pm 4 \pm 6\right)$\%
for \jpsi and \psitwos, and $\left(61 \pm 13\right)$\% for \chic.

The dominant systematic uncertainties are on the signal and background shapes
in the dimuon \pt fit, the muon identification efficiency, and the trigger
efficiency.

\section{Results}

The number of exclusive candidates passing the selection are 1492 \jpsi, 40
\psitwos, and 194 \chic, and the overall exclusive purity for $p_{\perp} <
900$~\mevc is estimated as $\left( 62 \pm 4 \pm 5 \right)$\% for the \jpsi,
$\left( 59 \pm 4 \pm 5 \right)$\% for the \psitwos, and $\left( 39 \pm 13
\right)$\% for the \chic.

The cross-section times branching fractions for central exclusive \jpsi,
\psitwos and \chic decaying to two muons with pseudorapidities between 2.0 and
4.5 are measured as
\begin{align*}
  \sigma_{\text{J}/\psi \rightarrow \mu \mu}
  \left( 2.0 < \eta_{\mu^{\pm}} < 4.5 \right) &= 307 \pm 21 \pm
  36\text{ pb,}\\
  \sigma_{\psi\left( 2 \text{S} \right)
    \rightarrow \mu \mu} \left( 2.0 < \eta_{\mu^{\pm}} < 4.5 \right)
  &= 7.8 \pm 1.3 \pm 1.0\text{ pb,}\\
  \sigma_{\chi_{\text{c}0} \rightarrow \text{J} / \psi \gamma
    \rightarrow \mu \mu \gamma} \left( 2.0 < \eta_{\mu^{\pm},\gamma} < 4.5 \right) &=
  9.3 \pm 2.2 \pm 3.5 \pm 1.8\text{ pb,}\\
  \sigma_{\chi_{\text{c}1} \rightarrow \text{J} / \psi \gamma
    \rightarrow \mu \mu \gamma} \left( 2.0 < \eta_{\mu^{\pm},\gamma} < 4.5 \right) &=
  16.4 \pm 5.3 \pm 5.8 \pm 3.2\text{ pb,}\\
  \sigma_{\chi_{\text{c}2} \rightarrow \text{J} / \psi \gamma
    \rightarrow \mu \mu \gamma} \left( 2.0 < \eta_{\mu^{\pm},\gamma} < 4.5 \right) &=
  28.0 \pm 5.4 \pm 9.7 \pm 5.4\text{ pb,}
\end{align*}
where the first uncertainty is statistical and the second systematic. The
measured \jpsi cross-section is consistent with theoretical predictions from
Gon\c{c}alves and Machado, Motyka and Watt, SuperCHIC and STARlight
\cite{Goncalves,Motyka,SuperCHIC,STARlight}.

\begin{figure}
  \centerline{\includegraphics[width=0.5\textwidth]{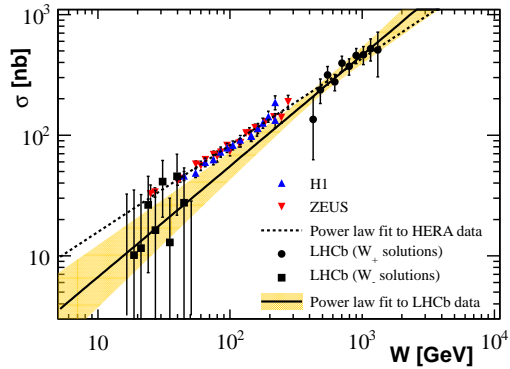}}
  \caption{The \jpsi photoproduction cross-section as a function of the
    photon-proton centre-of-mass energy. The red and blue points are HERA data,
    and the black points LHCb data. The dashed and solid lines are power law
    fits to the HERA and LHCb data respectively.}\label{fig:sigma_against_W}
\end{figure}

For comparison with HERA results, the differential \jpsi photoproduction
cross-section is measured in ten bins of \jpsi rapidity and reweighted by the
photon flux. There are two solutions in each rapidity bin, due to the ambiguity
over which proton emitted the photon. The two solutions are plotted in
Figure~\ref{fig:sigma_against_W}. With the limited precision of the
measurement, the LHCb results are consistent with HERA and confirm a similar
power law behaviour for the photoproduction cross-section \cite{H1,ZEUS}.

\section{Conclusion}

The cross-sections for central exclusive \jpsi and \chic production have been
measured in decays to dimuons with the LHCb detector, and the measurements
found to be consistent with a number of theoretical predictions. Additionally,
the exclusive \jpsi photoproduction cross-section has been measured as a
function of the photon-proton centre-of-mass energy. A power law fit shows
consistency with HERA results.

\bibliographystyle{abbrv}
\bibliography{bibliography}

\end{document}